\documentclass[aps,pra,showpacs,showkeys,twocolumn,superscriptaddress]{revtex4-1}
\usepackage[dvips]{color}
\usepackage{graphicx}
\usepackage{amssymb}

\begin{document}

%\begin{frontmatter}

\title{Semiclassics around a phase space caustic: an illustration using the Nelson Hamiltonian}

\author{A. D. Ribeiro}
\address{Departamento de F\'{\i}sica, 
Universidade Federal do Paran\'a, 81531-990, 
Curitiba, Paran\'a, Brazil}

\begin{abstract}
The semiclassical formula for the coherent-state propagator is written in terms of complex classical trajectories of an equivalent classical system. Depending on the parameters involved, more than one trajectory may contribute to the calculation. Eventually, however, two contributing trajectories coalesce, characterizing what is called phase space caustic. In this case, the usual semiclassical formula for the propagator diverges, so that a uniform approximation is required to avoid this singularity. In this paper, we present a non-trivial numerical application illustrating this scenario, showing the accuracy of the uniform formula that we have previously derived.
\end{abstract}

%\begin{keyword}
%semiclassical propagator \sep 
%coherent states \sep 
%phase space caustic \sep 
%uniform approximation
%\end{keyword}

%\end{frontmatter}

\maketitle

Quantum propagators are the fundamental ingredients in any dynamical description of the quantum theory. They also provide an important tool for the study of the quantum-classical connection since their semiclassical approximations can be intuitively interpreted in terms of classical trajectories. Naturally, therefore, they have been widely used in the context of the semiclassical theory. Concerned with the Correspondence Principle, Van Vleck~\cite{vv} inaugurated this kind of study by performing semiclassical approximations on the quantum propagator in the coordinate representation $\langle x''|\exp(-i\hat H T/\hbar)|x'\rangle$. According to his calculation, it can be written as a function of classical trajectories of the Hamiltonian $H_W$, the Weyl symbol of $\hat H$, connecting the initial position $x'$ to the final $x''$, after a time interval $T$. Around forty years later, this result was improved by Gutzwiller and used to derive his famous Trace Formula~\cite{gutzpaper,gutzbook}, which determines how periodic (and isolated) orbits of $H_W$ can be used to approximate the density of states of $\hat H$. Notice that Gutzwiller's work goes beyond a semiclassical description of dynamics. Actually, it filled a gap on semiclassical quantization methods since his formula applies to non-integrable systems, contrary to the earlier Bohr-Sommerfeld~\cite{BS} and Einstein-Brillouin-Keller~\cite{EBK1,EBK2,EBK3} quantization rules. 

Although these seminal developments on the semiclassical theory have essentially involved approximations in the coordinate representation, similar results can also be obtained by working with other representations. In particular, as classical states are usually points in phase-space, it is claimed to be natural to do semiclassical physics using the most localized quantum states in phase-space, requirement accomplished by the coherent states~\cite{cs,pere}. This natural predisposition of these states is corroborated by the great interest in time-evolution problems based on phase-space pictures, which can be appreciated, for instance, in Heller's papers~\cite{heller}. Moreover, the use of this representation has another advantage to be involved in semiclassical physics, namely, it can be easily extended in order to include spin degrees of freedom by means of spin coherent states~\cite{cs,pere}. This additional advantage can be identified already in the first paper that presents a semiclassical formula for the coherent-state propagator~\cite{z1d1}, since it considers both canonical and spin coherent states. At last, we recall that a derivation of the Gutzwiller Trace Formula using canonical coherent states can be found in Ref.~\cite{MC}, while the analog for spin coherent states can be found in Refs.~\cite{PZ,jmp}.

In this paper, we shall focus on semiclassical approximations of the two-dimensional coherent-state propagator 
\begin{equation}
 K(\mathbf z'', \mathbf z', T)=\langle \mathbf z''|\exp(-i\hat H T/\hbar)|\mathbf z'\rangle. 
\label{KQ}
\end{equation}
The states $|\mathbf z\rangle =|z_x\rangle\otimes|z_y\rangle$ are the coherent states that can be associated to a mass $m$ subjected to a harmonic potential with frequencies $\omega_{r}=\hbar/(mb^2_{r})$, with $r$ assuming $x$ or $y$. They are the eigenstates of the annihilation operator $\hat a_r$, namely, $ \hat a_r| z_r\rangle = z_r |z_r \rangle$, where
\begin{equation}
\hat a_{r} = \frac{1}{\sqrt2} 
\left(\frac{\hat q_{r}}{b_{r}}+i\frac{\hat p_{r}}{c_{r}}\right)
\;\; \mathrm{and} \;\;
z_{r} = \frac{1}{\sqrt2} 
\left(\frac{\bar q_{r}}{b_{r}}+i\frac{\bar p_{r}}{c_{r}}\right).
\end{equation}
Here, $c_r = \hbar/b_r$, $\hat q_r$ and $\hat p_r$ are, respectively, the position and momentum operators, $\bar q_r=\langle z_r|\hat q_r|z_r\rangle$, and $\bar p_r=\langle z_r|\hat p_r|z_r\rangle$. The numbers $b_r$ and $c_r$ can be also identified as the widths of $|\mathbf z\rangle$ in position and momentum, respectively. 

According to Refs.~\cite{ribeiro04,garg}, in the semiclassical limit, $K(\mathbf z'', \mathbf z', T)$ can be written in terms of functions depending only on trajectories of the classical Hamiltonian $H(\mathbf v,\mathbf u)$, which is achieved by calculating $\langle \mathbf z| \hat H|\mathbf z\rangle$, followed by the replacement of $\mathbf z$ and $\mathbf z^*$ by $\mathbf u$ and $\mathbf v$, respectively. The relation between the usual classical variables, $\mathbf q =(q_x,q_y)$ and $\mathbf p=(p_x,p_y)$, and the convenient variables, $\mathbf u=(u_x,u_y)$ and $\mathbf v=(v_x,v_y)$, is given by
\begin{equation}
u_{r} = \frac{1}{\sqrt2} 
\left( \frac{q_{r}}{b_{r}} +i\frac{p_{r}}{c_{r}}\right)
\;\; \mathrm{and} \;\;
v_{r} = \frac{1}{\sqrt2} 
\left( \frac{q_{r}}{b_{r}} -i\frac{p_{r}}{c_{r}}\right).
\label{nv}
\end{equation}
Hamilton's equations for $\mathbf u$ and $\mathbf v$ are
\begin{equation}
\dot u_r = -\frac{i}{\hbar} \frac{\partial H}{\partial v_r}
\quad \mathrm{and} \quad
\dot v_r = \frac{i}{\hbar} \frac{\partial H}{\partial u_r}.
\label{Keqm}
\end{equation}
The trajectories involved in the semiclassical evaluation of Eq.~(\ref{KQ}) must obey the boundary conditions
\begin{equation}
\mathbf u(0)=\mathbf z'
\quad \mathrm{and}\quad
\mathbf v(T)=\mathbf z''^*,
\label{bc1d} 
\end{equation}
which imply that $\mathbf q$ and $\mathbf p$ are complex, in general. Otherwise, both initial and final phase-space points would be fixed by the input, so that, generically, there would be no trajectory satisfying so many restrictions. This is the reason for the change $(\mathbf z^*,\mathbf z)\to (\mathbf v,\mathbf u)$. Once we have found such a contributing trajectory, we evaluate its complex action
\begin{equation}
\begin{array}{l}
\displaystyle
\mathcal{S}(\mathbf z''^*,\mathbf z',T) = 
\int_0^T 
\left[\frac{i\hbar}{2} 
\left(\dot\mathbf u\cdot\mathbf v-\mathbf u\cdot\dot\mathbf  v\right) 
- H
\right]dt 
-\Lambda,
\end{array}
\label{S}
\end{equation}
where $\Lambda = \frac{i\hbar}{2}\left[\mathbf u(0)\cdot\mathbf v(0)+\mathbf u(T)\cdot\mathbf v(T)\right]$, and 
\begin{equation}
\begin{array}{l}
\displaystyle
\mathcal{G}(\mathbf z''^*,\mathbf z',T) = 
\frac12 \int_0^T 
\left( \frac{\partial^2 H}{\partial u_x \partial v_x} +
\frac{\partial^2 H}{\partial u_y \partial v_y}\right)dt.
\end{array}
\label{G}
\end{equation}
The semiclassical propagator is then given by
\begin{equation}
\begin{array}{l}
K^{(2)}(\mathbf z''^*,\mathbf z',T) = 
\mathcal N ~{\displaystyle\sum_{\mathrm{traj.}}} 
\sqrt{ \det \left[\frac{i}{\hbar} 
\mathbf{S}_{\mathbf u \mathbf v} \right]}~ 
e^{\frac{i}{\hbar} \left(\mathcal S + \mathcal G \right)},
\end{array}
\label{K_original}
\end{equation}
where $\mathcal N = e^{-\frac{1}{2} |\mathbf z'|^2 -\frac{1}{2} |\mathbf z''|^2}$ and
\begin{equation}
\mathbf{S}_{\mathbf u \mathbf v} =
\left(\begin{array}{cc}
\frac{\partial^2\mathcal S}{\partial z'_x\partial z''^*_x} &  
\frac{\partial^2\mathcal S}{\partial z'_x\partial z''^*_y} 
\\
\frac{\partial^2\mathcal S}{\partial z'_y\partial z''^*_x}  &  
\frac{\partial^2\mathcal S}{\partial z'_y\partial z''^*_y}
\end{array}\right) .
\end{equation}
For non-integrable applications, it is convenient to write the prefactor $\mathcal P$ of Eq.~(\ref{K_original}) in terms of elements of~$\mathbf{M}$, which is the stability matrix of the contributing trajectory: $\mathcal P\equiv \sqrt{\det [(i/\hbar)\mathbf{S}_{\mathbf u \mathbf v} ]} = \sqrt{1/\det\mathbf{M_{vv}}} $, where
\begin{eqnarray}
\left(\begin{array}{l}
\delta \mathbf u(T) \\ \delta \mathbf v(T)
\end{array}\right) =
\left(\begin{array}{cc}
\mathbf{M_{uu}} & \mathbf{M_{uv}} \\
\mathbf{M_{vu}} & \mathbf{M_{vv}} 
\end{array}\right) 
\left(\begin{array}{l}
\delta \mathbf u(0) \\ \delta \mathbf v(0)
\end{array}\right).
\label{monodK}
\end{eqnarray}
Equation~(\ref{K_original}) is deduced by means of a quadratic approximation around critical paths (the complex classical trajectories) of $K(\mathbf z'',\mathbf z',T)$, written in the path integral formalism~\cite{ribeiro04,garg} (see also Refs.~\cite{z1d1,z1d4} for the one-dimensional case). This is the reason why we insert the index~$^{(2)}$ in the symbol $K$. In addition, it is explicitly indicated by the sum in Eq.~(\ref{K_original}) that, {\em in principle}, we should consider contributions of all trajectories satisfying boundary conditions~(\ref{bc1d}). At last, as trajectories depend just on $\mathbf z''^*$ instead of $\mathbf z''$, the label $\mathbf z''$ of $K$ is replaced by $\mathbf z''^*$ in ${K}^{(2)}$.

Some trajectories that obey Eqs.~(\ref{bc1d}), when used to calculate Eq.~(\ref{K_original}), give origin to non-physical results as, for instance, probabilities greater than one. This kind of problem has been reported in several papers~\cite{ribeiro04,nct1,nct2,nct3,nct4,nct5}, and it is assumed that these trajectories refer to critical points of the path integral quantum propagator impossible to be included in any allowed deformation of the original contour of integration. Usually, these trajectories are simply excluded from the calculation. A simple and useful rule to identify such spurious trajectories consists in writing their contributions to $K^{(2)}$ as $e^{iF/\hbar}$, so that we can select the ones whose imaginary part of $F_0$, defined as the zero-order term of $F$ in its $\hbar$-expansion, is non-negative. Otherwise, in the formal semiclassical limit $\hbar\to0$, their contributions ($\sim e^{-\mathrm{Im}[F_0]/\hbar}$) would produce a non-physical $|K^{(2)}|$. Notice that, although $\mathcal S$ may have terms on $\hbar$, it is a good estimate to think of $\mathrm{Im}[F_0]= \mathrm{Im}[\mathcal S] - \hbar \ln\mathcal N \equiv \mathcal F_0$ (for a careful discussion about the $\hbar$-dependence of each term of Eq.~(\ref{K_original}), see Ref.~\cite{z1d4}). 

It should be mentioned that the abrupt removal of a contribution from Eq.~(\ref{K_original}) is a manifestation of the well-known Stokes Phenomenon~\cite{langer,sp1,berry89}. Generically, it appears when an analytic function (Eq.~(\ref{KQ}), in our case) is asymptotically ($\hbar\to0$) approximated by a multi-valued function [the sum of exponential contributions~(\ref{K_original})]. Stokes Phenomenon refers to the fact that the proper choice of a branch in the approximating function is domain-dependent. We emphasize that the sudden change in the {\em form} of the approximating function does not represent its (numerical) discontinuity. Actually, it is needed in order to assure the continuity manifested in the function represented. The criterion concerning the sign of $\mathcal F_0$ presented earlier combined with considerations on continuity shall be, therefore, our basis to decide if a trajectory should contribute to the propagator or not. Finally, as a generic manifestation of asymptotic approximations, the phenomenon is quite often in semiclassical physics. Apart from the cases cited earlier concerning the coherent-state propagator, it can be observed, for instance, in the WKB method~\cite{langer,maslov1} and also in the propagator in the momentum representation~\cite{shudo}.

Besides the problem of spurious contributions in Eq.~(\ref{K_original}), it may appear trajectories for the which the prefactor $\mathcal P$ diverges. The point where it happens is called phase space caustic (PSC), and it is caused by the coalescence of contributing trajectories. From the mathematical point of view, it arises because second order corrections of the expansion of $K(\mathbf z'',\mathbf z',T)$ around the classical trajectory vanish. To avoid this problem, we need to develop improved approximations where further corrections are considered. As it arises from the approach performed and not because of the trajectory itself, we point out that, in this case, there is no reason to exclude a trajectory from the calculation. We shall return to the treatment of this issue opportunely. 

%%%%%%%%%%%%%%%%%%%%%%%%%%%%%%%%%%%%%%%%%%%%
\begin{figure*}
\includegraphics[width=9cm,angle=-90]{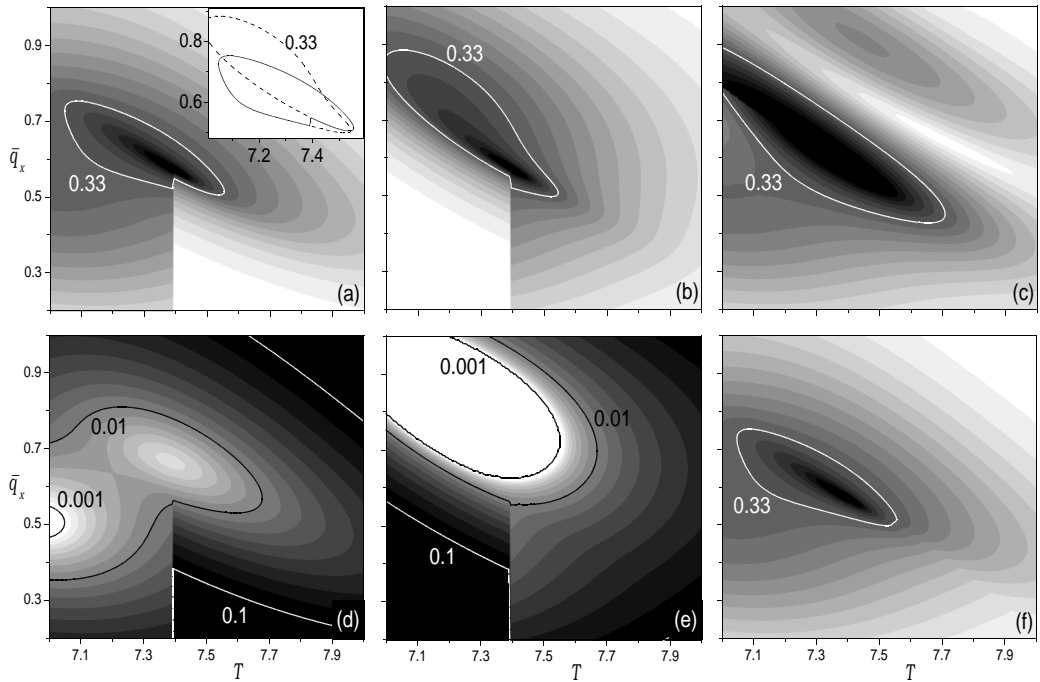}
\caption{Panels (a)-(c) show the contour plots of $|K^{(2)}|$ in the $(T,\bar q_x)$ plane. The individual contribution of family $f_a$ is shown in panel (a), while the one of family $f_b$ is shown in panel (b). The inset of panel (a) shows the contour curve $|K^{(2)}|=0.33$ for families $f_a$ (solid line) and $f_b$ (dashed line) superimposed. Panel (c) presents $|K^{(2)}|$ evaluated with both families. Panels (d) and (e) show the contour plot of $\mathcal F_0$ for families $f_a$ and $f_b$, respectively. While $f_a$ has no point where $\mathcal F_0<0$, for $f_b$ practically the whole region inside the contour curve $\mathcal F_0=10^{-3}$ has $\mathcal F_0<0$. Panel (f) combines the plots (a) and (b). In all contour plots, including those of Fig.~\ref{fig2}, some contours were highlighted to facilitate a comparison among them. For panels (a)-(c) and (f), the difference between two subsequent curves is 0.03 [the same for Fig.~\ref{fig2}(a)-Fig.~\ref{fig2}(d)]. Axes are the same for all graphs.}
\label{fig1}
\end{figure*}
%%%%%%%%%%%%%%%%%%%%%%%%%%%%%%%%%%%%%%%%%%%%%%

In order to illustrate this rich scenario of spurious trajectories and PSC's, we apply Eq.~(\ref{K_original}) to
\begin{equation}
\begin{array}{l}
\displaystyle
\hat H = \frac12 \left( \hat p_x^2+\hat p_y^2\right) + 
\left(\hat q_y - \frac{\hat q_x}{2}\right)^2 + 
\frac{\mu}{2} \hat q_x^2,
\end{array}
\label{NelsonO}
\end{equation}
known as Nelson Hamiltonian, which has been studied in both Classical~\cite{tc1,tc2} and Quantum Mechanics~\cite{tq1,tq2}. Actually, in Ref.~\cite{ribeiro04}, we had already used this system to study the applicability of Eq.~(\ref{K_original}). Now, we revisit this work in order to deal with the problem of PSC there presented, but not solved. As well as we previously did~\cite{ribeiro04}, we shall restrict the application to the case where $\mathbf z'=\mathbf z''= \mathbf z$. In addition, we shall also use $b_x=b_y=0.2$, $\mu=0.1$ and $\hbar=0.05$. By doing so, five numbers ($\bar q_x$, $\bar q_y$, $\bar p_x$, $\bar p_y$, and $T$) become the input parameters to calculate $K^{(2)}$. We then define $\bar p_x=|\bar \mathbf p|\cos\theta$, $\bar p_y=|\bar \mathbf p|\sin\theta$, and
\begin{equation}
\begin{array}{l}
\displaystyle
E = \frac12 |\bar \mathbf p|^2 + 
\left(\bar q_y - \frac{\bar q_x}{2}\right)^2 + \frac{\mu}{2} \bar q_x^2,
\end{array}
\label{energy}
\end{equation}
holding $E=0.5$ and $\theta=140^\circ$. Thus, for a given input pair $(T,\bar q_x)$, we select $\bar q_y$ by the rule $\bar q_y=2\bar q_x/3$, so that the last undetermined parameter $\bar \mathbf p$ is solved by the last equation. We point out that this set of parameters was chosen in order to find a region containing a PSC, but described by a reduced number of variables, namely, $T$ and $\bar q_x$. In the following, we show the evaluation of Eq.~(\ref{K_original}) in the plane $(T,\bar q_x)$ for the interval $0.2<\bar q_x<1.0$ and $7.0<T<8.0$.

\begin{figure*}
\includegraphics[width=9cm,angle=-90]{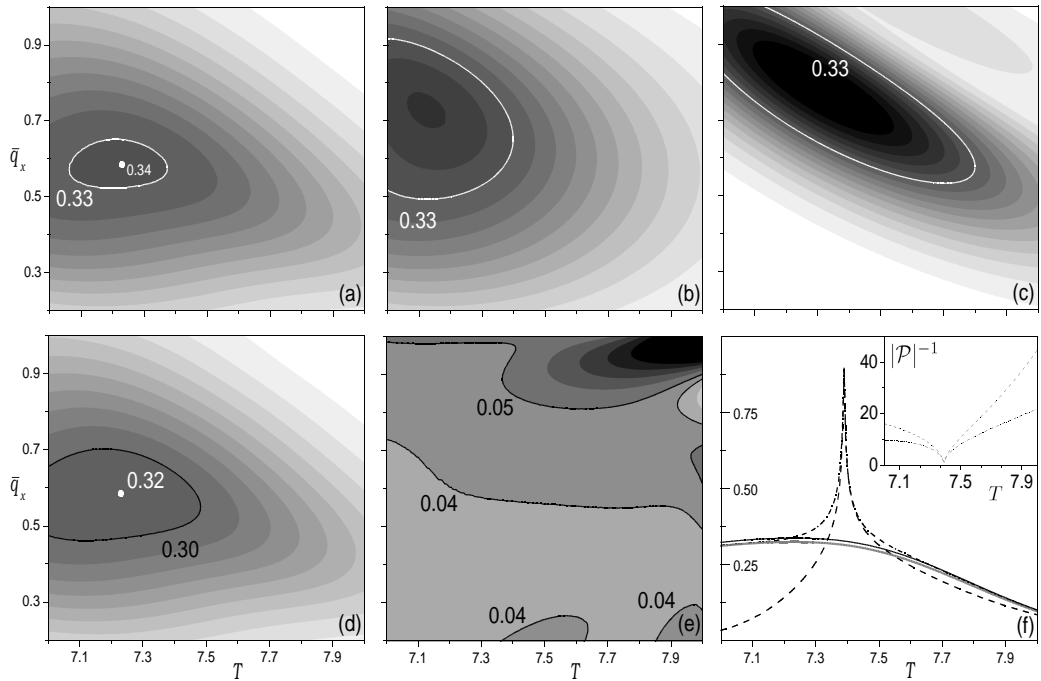}
\caption{Panels (a)-(c) show three continuous and distinct contour plots for the solutions of $|K^{\mathrm{(un)}}|$ in the $(T,\bar q_x)$ plane. Panel~(d) shows the contour plot of the exact $|K|$, and panel (e) the relative error between panels (a) and (d). In panel~(f), we present, just for $\bar q_x=0.58$, the results of Fig.~\ref{fig2}(a) (gray solid line), Fig.~\ref{fig2}(d) (black solid line),  Fig.~\ref{fig1}(a) (dash-dotted line), and Fig.~\ref{fig1}(b) (dashed line); Its inset shows $|\mathcal P|^{-1}$ for $f_a$ (dash-dotted line) and $f_b$ (dashed line), for the same points. For panel~(e), the difference between two subsequent curves is 0.01. Axes are the same for all graphs, except for panel (f) where they are explicitly shown.}
\label{fig2}
\end{figure*}

For all points $(T,\bar q_x)$ considered, we found two contributing trajectories to $K^{(2)}$~\cite{comm}. Based on continuity criteria, we can distinguish two families of such trajectories, $f_a$ and $f_b$. The individual contribution of each family to Eq.~(\ref{K_original}) is shown in Fig.~\ref{fig1}(a), for $f_a$, and Fig.~\ref{fig1}(b), for $f_b$. Although a clear vertical cut line appears in the plots, we point out that the combination of both families gives origin to a continuous two-branch surface exhibiting a M\"obius strip structure. In the inset of Fig.~\ref{fig1}(a), we demonstrate this property: By circulating the contour curve $|K^{(2)}|=0.33$ of $f_a$, to avoid the discontinuity at the cut line, one should change to the contour curve $|K^{(2)}|=0.33$ of $f_b$. Then, if one continues to follow this curve, one arrives again at the cut line, where one can continuously return to family $f_a$, closing a cycle of two turns. In Fig.~\ref{fig1}(c), we evaluate $|K^{(2)}|$ including both families. Notice that these three plots present a sharp peak that demands investigation. In Fig.~\ref{fig2}(f), therefore, we plot the results of Figs.~\ref{fig1}(a) and \ref{fig1}(b) just for the line $\bar q_x=0.58$, where we clearly identify a divergent behavior. As shown in the inset of this figure, it appears because of the presence of a PSC, point where $\mathcal P^{-1}$ goes to zero.

Apart from the region under influence of the PSC, which can not be properly evaluated by Eq.~(\ref{K_original}), the question that naturally arises is about which plot satisfactorily approaches the equivalent full quantum mechanical calculation. In order to answer this point, we plot in Fig.~\ref{fig1}(d) and Fig.~\ref{fig1}(e) the value of $\mathcal F_0$ for $f_a$ and $f_b$, respectively. From them, and according to the criterion concerning the sign of $\mathcal F_0$ defined above, we conclude that there is no reason why to exclude $f_a$ from Eq.~(\ref{K_original}). On the other hand, for $f_b$, there is a large region where $\mathcal F_0<0$, implying that this family can not be used to evaluate $|K^{(2)}|$ in this region. Then, if we assume that {\em only} $f_a$ contributes to this particular region and impose continuity in the whole plane $(T,\bar q_x)$, we find the result shown in Fig.~\ref{fig1}(f). In practice, to plot this graph, except for the region where we know that $f_b$ should not be included, we span the whole plane $(T,\bar q_x)$, comparing the results of Figs.~\ref{fig1}(a) and \ref{fig1}(c) and selecting the one which optimizes continuity. Figure~\ref{fig1}(f), in this sense, is the better we can do by using Eq.~(\ref{K_original}). A close look at Figs.~\ref{fig1}(a), \ref{fig1}(c), \ref{fig1}(e), and \ref{fig1}(f), however, reveals that family $f_b$ was excluded from the calculation even where $\mathcal F_0$ is non-negative. It could be seem as an illegitimate procedure, but, as discussed earlier in the present paper, supported by the Stokes Phenomenon, we are allowed to do this. 

%%%%%%%%%%%%%%%%%%%%%%%%%%%%%%%%%%%%%%%%%%%%%%
%%%%%%%%%%%%%%%%%%%%%%%%%%%%%%%%%%%%%%%%%%%%%%

We now face the problem of the PSC. There is no other solution to this issue unless to revisit and to improve the approximations performed in Eq.~(\ref{KQ}). Two ingredients are crucial to accomplish this task: Maslov's method~\cite{maslov2,berry83,ozo87} and uniform approximations~\cite{langer,kus93,un00,un01,un02,bleistein}. Generically, the first one consists in working with two conjugate representations of the same semiclassical object, so that if there is a singularity in a given representation one changes to other. Then, by transforming back to the original representation including further corrections, the singular point can be circumvented. The method has the advantage of avoiding the extremely complicated calculation originated by the direct implementation of third order corrections in the path integral representation of Eq.~(\ref{KQ}). Uniform approximations, in this case, are useful to perform the integral involved in the last step of the Maslov method. They enable us to map the complicated integrand into a simpler one, but having a similar structure of saddle points. Essentially, as in the asymptotic limit $\hbar\to0$ saddle points concentrate useful information about the integral, the mapping assures a good accuracy. We recall that the combination of both methods were already used to deal with problems similar to PSC's, namely, to deal with the break-down of quadratic approximations. It was used, for instance, to avoid the turning point divergence in the WKB method~\cite{maslov1,berry83} and also to treat caustics in the coordinate propagator~\cite{sieber97}. In Ref.~\cite{sieber97}, in particular, Schomerus and Sieber derived extensions of the Gutzwiller formula for the case of coalescent orbits. We point out that, while in Ref.~\cite{sieber97} the methods were applied to perform the trace and the Fourier transform of the coordinate propagator (operations which transform the propagator into the uniform density of states), our approximation is exclusively performed in order to find a uniform formula for the propagator itself. 

Following these ideas, we studied a conjugate representation for the coherent states~\cite{jpa2009}. From it the Maslov method can be applied so that a uniform approximation for $K$, valid for regions close and far from PSC's, can be easily obtained. This task was already performed for both 1D~\cite{un1} and 2D coherent states~\cite{un2}, with some simple applications presented in Ref.~\cite{jpcs}. The prescription that we achieved to properly evaluate the propagator around a PSC includes, firstly, finding the two contributing trajectories to $K^{(2)}$. Then, from their actions $\mathcal S_a$ and $\mathcal S_b$, we calculate
\begin{equation}
\begin{array}{l}
\mathcal A = \frac{i}{2\hbar}(\mathcal S_a + \mathcal S_b)
\quad \mathrm{and}\quad
\mathcal B = 
\left[\frac{3i}{4\hbar}(\mathcal S_b - \mathcal S_a)\right]^{2/3}, 
\end{array}
\label{AB}
\end{equation}
which can be directly used in the uniform formula~\cite{jpcs}
\begin{equation}
K^{(\mathrm{un})}(\mathbf z''^*,\mathbf z',T) = 
i\sqrt\pi
\left[c_1
\mathrm{f}_j'(\mathcal B)+ c_2
\mathrm{f}_j(\mathcal B)
\right]e^{\mathcal A},
\label{K_un}
\end{equation}
where $c_1 = (h_b-h_a)/\sqrt\mathcal B$ and $c_2=h_a+h_b$, with
\begin{equation}
h_{a,b} = \sqrt{\mp \sqrt\mathcal B
/(\det \mathrm{M_{vv}})|_{{a,b}}}~
e^{\frac{i}{\hbar} \mathcal G_{a,b}}.
\end{equation}
The function $\mathrm{f}_j(\xi)$ is the well-known Airy's function,
\begin{equation}
\begin{array}{l}
\displaystyle
\mathrm{f}_j(\xi)=\frac{1}{2\pi}\int_{C_j} 
\exp\left\{i\left(\xi t+\frac{t^3}{3}\right)\right\}dt,
\end{array}
\label{airy}
\end{equation}
where the index $j$ refers to three possible paths of integration $C_j$, related to three different Airy's functions~\cite{bleistein}.

When used for the present application, the three possible solutions for the uniform formula~(\ref{K_un}) can be easily organized as three continuous and distinct solutions in the plane $(T,\bar q_x)$, as shown in Figs.~\ref{fig2}(a), \ref{fig2}(b) and~\ref{fig2}(c). Comparing them with the results of Fig.~\ref{fig1}, we realize that Fig.~\ref{fig2}(a), Fig.~\ref{fig2}(b), and Fig.~\ref{fig2}(c) refer, respectively, to the uniformization of Fig.~\ref{fig1}(a), Fig.~\ref{fig1}(b), and Fig.~\ref{fig1}(c). Since Fig.~\ref{fig2}(a) agrees with Fig.~\ref{fig1}(f) far from the PSC, we elect it as the final semiclassical result. In Fig.~\ref{fig2}(d), we show the full quantum mechanical result, and, in Fig.~\ref{fig2}(e), the relative error $||K|-|K^{(\mathrm{un})}||/|K|$ between Fig.~\ref{fig2}(a) and Fig.~\ref{fig2}(d). Notice that the adopted approximation satisfactorily agrees with the exact result: For almost all points, the error is less than 5\%. At last, for an additional comparison, in Fig.~\ref{fig2}(f), we plot in the same graph the two individual contributions for $K^{(2)}$, the exact, and the uniform result of Fig.~\ref{fig1}(a), just for the line $\bar q_x=0.58$. We emphasize that the choice of the branch of Eq.~(\ref{K_un}) can also be seem as a manifestation of the Stokes Phenomenon. Notice also that, once the singularity has been removed, continuity is more evident and the choice becomes easier.

In this article, we briefly reported a rich numerical study on semiclassical approximations of the coherent-state propagator. Analogously to the most famous semiclassical approximation based on second order expansion, namely, the WKB formula, $K^{(2)}$ also suffers the problem of non-physical solutions and singularities in its prefactor. The first problem can be handled by excluding spurious trajectories using continuity (physical) criteria. This elimination, however, does not solve the problem of PSC (equivalent to the turning point divergence in the WKB method). Combining the Maslov method with a conjugate for the coherent-state representation, we developed a uniform formula for $K$, which has shown to agree with the full quantum mechanical calculation. Curiously, in the uniform approach, spurious trajectories of $K^{(2)}$ become crucial to calculate $K^{(\mathrm{un})}$. Concerning the decision about which solution of $K^{(\mathrm{un})}$ should be chosen, it can be done by imposing continuity and the fact that $K^{(\mathrm{un})}$ should agree with $K^{(2)}$ in regions far from PSC's. Finally, we emphasize that the numerical example studied in the present paper figures among the worst scenarios to evaluate semiclassical formulas, since we approach very close to a caustic. In spite of the adverse conditions, the systematic use of techniques concerned with both standard second order approximation and uniform approximation has shown to be quite satisfactory to deal with this situation.

This work was supported by CNPq (474096/2008-4). The author wishes to thank M.~A.~M. de Aguiar for helpful discussions.

\bibliographystyle{elsarticle-num}

\begin{thebibliography}{20}
\bibitem{comm} In Ref.~\cite{ribeiro04}, for the same calculation, we had not found two trajectories for all points considered. However, we returned to the problem and improved the search algorithm, so that now we have two trajectories per point.
\end{thebibliography}

\end{document}